\documentclass[12pt]{iopart}

\usepackage{graphicx}
\usepackage{textcomp}
\usepackage{color}
\usepackage{dcolumn}
\usepackage{bm}

\newcommand{\ket}[1]{\mbox{$ | #1 \rangle $}}
\newcommand{\equationname}[1]{\textsc{Eq.}~#1}

\begin{document}

\title[Polarization entanglement preparation and manipulation% in standard telecommunication channels
]{High-quality polarization entanglement state preparation and manipulation in standard telecommunication channels}

\author{F Kaiser$^{1}$, A Issautier$^{1}$, L A Ngah$^{1}$, O D\u{a}nil\u{a}$^{1,2}$, H Herrmann$^3$, W Sohler$^3$, A Martin$^{1,\star}$ and S Tanzilli$^{1}$}%
\address{$^1$Laboratoire de Physique de la Mati\`ere Condens\'ee, CNRS UMR 7336,
Universit\'e de Nice -- Sophia Antipolis, Parc Valrose, 06108 Nice Cedex 2, France.}
\address{$^2$Faculty of Electronics, Communication and Information Technology, University Politehnica of Bucharest, Blvd. Iuliu Maniu 1-3, 061071 Bucharest, Romania}
\address{$^3$Applied Physics, University of Paderborn, 33098 Paderborn, Germany}
\address{$^\star$Currently with the Group of Applied Physics, University of Geneva, Switzerland}
\ead{sebastien.tanzilli@unice.fr}

\begin{abstract}
We report a simple and practical approach for generating high-quality polarization entanglement in a fully guided-wave fashion. Both deterministic pair separation into two adjacent telecommunication channels and the paired photons' temporal walk-off compensation are achieved using standard fiber components. Two-photon interference experiments are performed, both for quantitatively demonstrating the relevance of our approach, and for manipulating the produced state between bosonic and fermionic symmetries. The compactness, versatility, and reliability of this configuration makes it a potential candidate for quantum communication applications.
\end{abstract}


\maketitle
\tableofcontents

\section{Introduction}

Photonic entanglement is a key resource for quantum communication protocols~\cite{Tittel_photonic_2001}, among which today's main application is quantum key distribution~\cite{Gisin_QKD_2002,Scarani_QKD_2009}.
Producing entangled photons at telecom C-band wavelengths (1530-1565\,nm) enables implementing relatively long-distance quantum links, thanks to low-loss optical fibers and high performance standard components. One of the most natural and accessible entanglement observables is the polarization state of the photons. Many experiments have been performed in this field, often taking advantage of the so-called type-II spontaneous parametric down conversion (SPDC) process in suitable non-linear bulk~\cite{Kwiat_source_1995,Piro_SPSA_2009} or waveguide~\cite{Zhong_typeIIsecond_2010,Martin_typeIIsecond_2010} crystals. %, such as LiNbO$_3$ and KTP. 
Moreover, integrated optics in LiNbO$_3$ is a powerful resource for engineering quantum circuits for both generating and manipulating entangled photons~\cite{Tanz_LPR_2012}. Strong nonlinear optical effects as well as single spatial mode operation are indeed key ingredients for future developments of quantum information technology~\cite{Martin_IQR_2012}.

In this paper we present a simple and practical approach for generating high-quality polarization entanglement in a fully guided-wave fashion. Both deterministic pair separation in two adjacent telecommunication channels and walk-off compensation are achieved using standard fiber components. Two-photon interference experiments are carried out both for precisely determining the fiber compensation stage, and for manipulating the produced state between bosonic ($\ket{\Psi^+}$) and fermionic ($\ket{\Psi^-}$) symmetries~\cite{Matthews_BosonFermion_2011}. Eventually, we perform a Bell inequality test on the $\ket{\Psi^+}$ state to quantitatively demonstrate the relevance of both our approach and results.

In the case of type-II waveguides, paired photons are emitted co-propagatively in the cross-polarization states $\ket H \ket V$. Successful formation of entanglement is then obtained upon probabilistic pair separation at a beam-splitter, at the price of a 50\% pair loss. To circumvent this, several strategies have been proposed. They are based on either multi-waveguide configurations~\cite{Yoshizawa_twoPPLN_2003,Jiang_Sagnac_2007,Kawashima_typeI_2009}, interlaced poling periods in lithium niobate~\cite{Kryshna_doublepoling_2009,Suhara_quasiphase_2009,Thomas_interlaced_2010}, or concurrent type-0/I SPDC processes in AlGaAs ridge waveguides~\cite{Kang_concurrent_2012}. Related experiments reported limited entanglement qualities due to technological imperfections~\cite{Yoshizawa_twoPPLN_2003,Jiang_Sagnac_2007,Kawashima_typeI_2009,Suhara_quasiphase_2009,Thomas_interlaced_2010}. In addition, the temporal walk-off between the paired photons naturally induced by the waveguide birefringence needs to be compensated. In earlier works, the latter was achieved using bulk optics strategies, reducing both the reliability and the applicability of the source~\cite{Zhong_typeIIsecond_2010,Martin_typeIIsecond_2010}. Here, we address these two issues using a low-loss telecom fiber solution.

\section{Principle of the source for deterministic paired-photons separation}

\begin{figure}[h!]
\includegraphics[width=0.9\columnwidth]{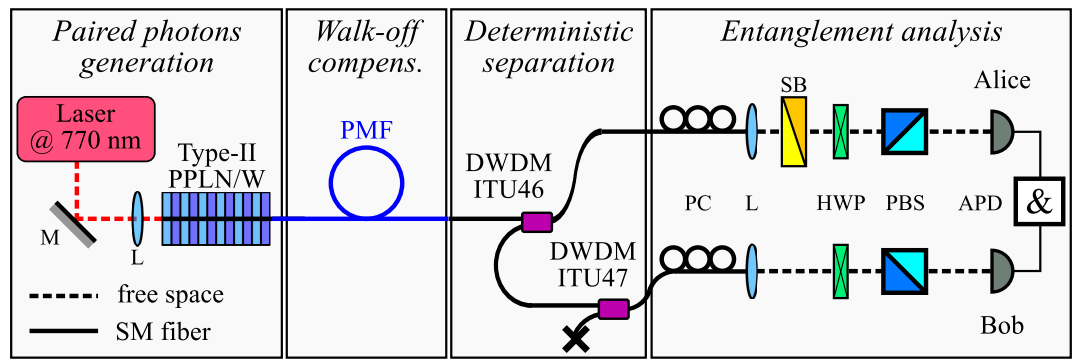}
\caption{Setup for generating polarization entangled photons at 1540\,nm using a type-II PPLN/W pumped by a 770\,nm laser. A PMF is used as birefringence compensator and a set of two DWDMs separates the photons deterministically into the ITU-46/47 channels. A standard Bell inequality apparatus measures the entanglement quality. On Alice's side, a Soleil-Babinet phase compensator (SB) permits adjusting the phase of the entangled state. L: lens; PC: fiber polarization controller; HWP: half wave-plate; PBS: polarising beam-splitter; \&: AND-gate.
\label{Fig1_Fullsetup}}
\end{figure}

\begin{figure}[h!]
\includegraphics[width=0.6\columnwidth]{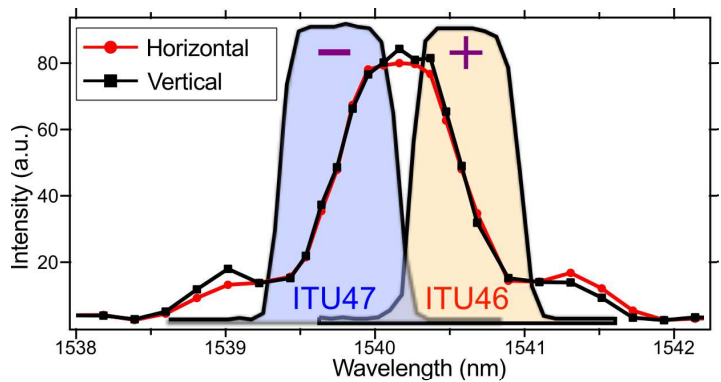}
\caption{Polarization dependent emission spectrum of the PPLN/W at 110$^{\circ}$C. The two polarization modes show a near-perfect overlap. Filled curves represent the transmission spectra of the two employed DWDM filters. $+$ and $-$ signs indicate the high and low coupled wavelength transmission windows, respectively.\label{Fig2_Spectrum}}
\end{figure}

The experimental setup of the source is shown in \figurename~\ref{Fig1_Fullsetup}.
The type-II photon-pair generator is based on a 3.6\,cm long, 9.0\,$\mu$m periodically poled, lithium niobate waveguide structure (PPLN/W)~\cite{Martin_typeIIsecond_2010}. More details of our fabrication processes and numerical simulations carried out for reaching the desired phase-matching are given in~\cite{martin_integrated_2009}. We couple 2.5\,mW (nominal power) from a 770\,nm diode laser into the waveguide towards creating paired photons around the degenerate wavelength of 1540.2\,nm.
As shown in \figurename~\ref{Fig2_Spectrum}, the $\ket H$ and $\ket V$ emission modes overlap almost perfectly with a spectral bandwidth of about 0.85\,nm FWHM.

We address the paired photons deterministic splitting via a pertinent strategy which consists of dividing the emission spectrum of \figurename~\ref{Fig2_Spectrum} into high ($+$) and low ($-$) wavelength windows, so as to create a state of the form
\begin{equation}
\ket{\Psi(\phi)} = \frac{1}{\sqrt{2}} \left ( \ket H_{+} \ket V_{-} + {\rm e}^{{\rm i}\,\phi} \ket V_{+} \ket H_{-} \right).
\label{Eq_Psi}
\end{equation}
To this end, we employ two complementary standard 100\,GHz dense wavelength division multiplexers (DWDM). As shown in \figurename~\ref{Fig2_Spectrum}, this set of filters is chosen so as to nearly cover the full emission spectrum, but also to remove the side peaks of the spectrum that provide with a polarization determination as a function of the wavelength~\cite{Martin_typeIIsecond_2010}. This way, almost no useful photons are lost in the spectrum central peak. Note that the actual overlap between the two filters at the center of the spectra is on the order of 1\%. Taking into account the relative spectral intensities, this leads to a probability of $<$0.25\% of having the two photons sent to the same wavelength region, \textit{i.e.} $\ket V_{+} \ket H_{+}$ and $\ket V_{-} \ket H_{-}$ given to the same user ($+$ to Alice and $-$ to Bob, respectively).
Therefore, additional noise contributions, due to the detection of a non-entangled photon in one wavelength region and a dark-count in the other detector, are negligible.

\section{Temporal walk-off compensation using a fiber approach}

\begin{figure}[h!]
\includegraphics[width=0.8\columnwidth]{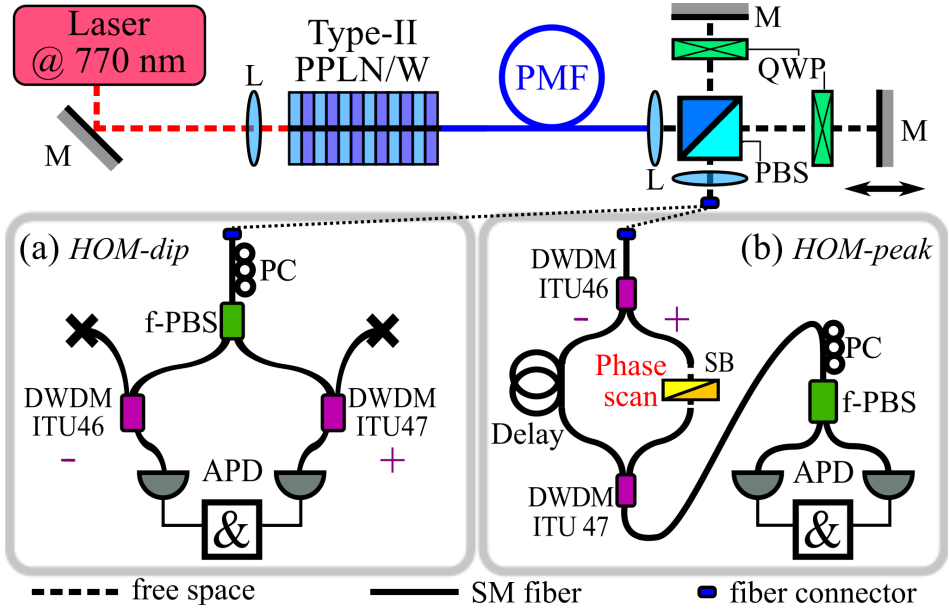}
\caption{Experimental setup for (a) observing the HOM-dips with and without the PMF, and (b) shifting the phase of the entangled state. In the former situation, the photons states are rotated by 45$^\circ$ thanks to a fiber PC and sent to an f-PBS for state analysis. Wavelength post-selection is made using two DWDMs centred in the ITU-46 and 47 channels. In the latter situation, the paired photons are deterministically separated at a first DWDM (ITU-46). Photons in the $-$ region are subjected to a variable delay, while those in the $+$ region are subjected to a polarization dependent phase shift introduced thanks to an SB plate. Eventually, the paired photons are recombined at a second DWDM (ITU-47), rotated by 45$^\circ$ (PC), and sent to an f-PBS for state analysis. L: lens; QWP: quarter wave-plate; M: Mirror; \&: AND-gate.\label{Fig3_Setup_DipPeak}}
\end{figure}

To compensate the temporal walk-off between the $\ket{H}$ and $\ket{V}$ photons induced by the waveguide birefringence, we take advantage of a simple, reliable, and low-loss strategy. As previously used with fiber-based photon-pair sources~\cite{Li_PMF_2005,Li_Storage_2005}, a birefringent polarization maintaining fiber (PMF) is butt-coupled to the waveguide's output and oriented such that the fiber birefringence axis is rotated by 90$^{\circ}$ compared to that of the waveguide.
To find the correct compensation length of the PMF, we perform a two-photon interference experiment based on the so-called Hong, Ou, and Mandel (HOM) dip setup~\cite{HOM_dip_1987} for the polarization observable~\cite{Martin_typeIIsecond_2010}, in which the temporal delay between the two photons is scanned. On one hand, this permits inferring the natural temporal walk-off induced by the waveguide birefringence. On the other hand, it allows characterizing the specific walk-off compensation achieved per meter of PMF.
The HOM-like experimental apparatus is shown in \figurename~\ref{Fig3_Setup_DipPeak} with the left inset (a) connected to the output of the source.

The generated photons are first sent to a tunable polarization dependent delay based on a Michelson interferometer like arrangement, as detailed in~\cite{Martin_typeIIsecond_2010}. One of the mirrors can be displaced using a motor to introduce an artificial delay between the photons. Thereafter, the bi-photon state is rotated by 45$^{\circ}$ using a fiber polarization controller (PC) compared to the $\{H,V\}$ creation basis, and sent to a fiber polarizing beam-splitter (f-PBS). After the f-PBS, two DWDMs are used for selecting high ($+$) and low ($-$) wavelength photons. Note that in principle, the actual position of the filtering stage, \textit{i.e.} before or after the f-PBS, does not matter. This allows considering the state of \equationname{~\ref{Eq_Psi}}, with $\phi$=0, at the input of the measurement apparatus of \figurename~\ref{Fig3_Setup_DipPeak}(a). In this case, the evolution of the sate reads
\begin{equation}
\begin{array}{l c}
%\hspace{-2cm}
\ket{\Psi} = \frac{1}{\sqrt{2}} \left ( \ket H_{+} \ket V_{-} + \ket V_{+} \ket H_{-} \right )\vspace{0.1cm}\\
%\hspace{-1cm}
\stackrel{45^{\circ}}{\longmapsto} \frac{1}{2\sqrt{2}} \left ( \ket H_+ \ket V_- - \ket H_+ \ket H_- + \ket V_+ \ket V_- - \ket V_+ \ket H_- \right. \\
\begin{array}{c c}
\hspace{1cm}\, & \left. + \ket V_+ \ket H_- + \ket V_+ \ket V_- - \ket H_+ \ket H_- - \ket H_+ \ket V_- \right ).
\end{array}
\end{array}
\end{equation}
When the delay between the photons is perfectly compensated, the states associated with cross-polarized contributions cancel each other. As a consequence, the quantum state after the action of both the f-PBS and the filters leads to the state
\begin{equation}
\ket{\Psi_{\rm HOM}} = \frac{1}{\sqrt{2}} \left ( \ket V_{+} \ket V_{-} - \ket H_{+} \ket H_{-} \right).
\end{equation}

\begin{figure}[h!]
\includegraphics[width=0.7\columnwidth]{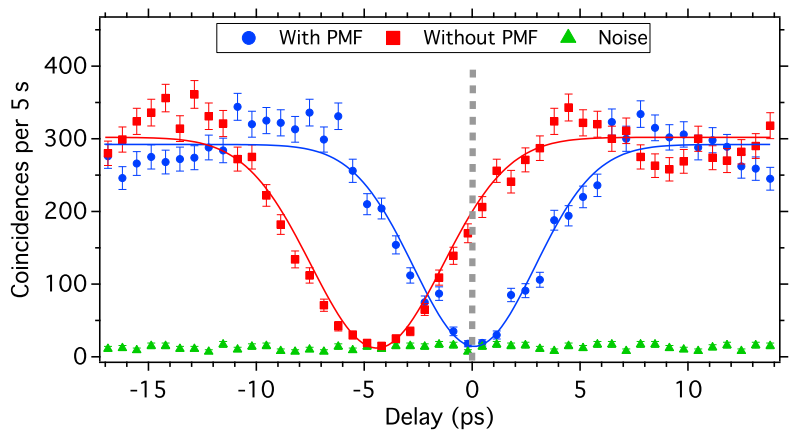}
\caption{HOM-dip measurements with and without the PMF. The vertical dashed line symbolizes zero delay.\label{Fig4_HOMdips}}
\end{figure}

The coincidence detection rate is then recorded, as a function of the artificial delay, between two InGaAs avalanche photodiodes (APD). One is a passively quenched device (IDQ-220, featuring $\sim$20\% detection efficiency and $\sim$10$^{-6}$\,/\,ns dark-count probability) which triggers, upon detection, the second one operated in gated mode (IDQ-201, featuring $\sim$25\% detection efficiency and $\sim$10$^{-5}$\,/\,ns dark-count probability). As shown in \figurename~\ref{Fig4_HOMdips}, we obtain HOM-dips when performing the experiment with and without the PMF walk-off compensator. The associated raw visibilities are both of 94$\pm$5\%. Note that no particular efforts have been made towards optimizing these figures. Without the PMF, the dip is centred at a delay of $-4.40\rm\,ps$, which corresponds to the natural walk-off induced by the waveguide birefringence, as predicted by the theory~\cite{Martin_typeIIsecond_2010}. We characterized the specific PMF compensation to be $\rm 1.38\,ps/m$, meaning that a 3.2\,m long PMF is necessary in our case. With such a fiber, the dip position is shifted to 0.03\,ps, indicating that the walk-off is nearly perfectly compensated.

\section{Manipulation of the entangled state symmetry}

The above dips have been obtained with photons at different wavelengths, contrary to what is commonly done in similar HOM measurements~\cite{Zhong_typeIIsecond_2010,Martin_typeIIsecond_2010}. This indicates that this quantum effect does not simply originate from a conventional two-photon interference, but rather from an entangled state post-selected by the measurement apparatus (f-PBS+DWDMs).
Namely, in standard polarization HOM experiments, \textit{i.e.} involving wavelength degenerate photons, the dip is observed independently of potential phase fluctuations between the two photons impinging the PBS~\cite{Martin_typeIIsecond_2010}. Entanglement, however, implies a constant phase relation between its two contributions.
To show this, we now discuss a second measurement which makes it possible to switch the phase of the entangled state, if any, from $0$ to $\pi$, corresponding to the $\ket{\Psi^+}$ and $\ket{\Psi^-}$ maximally entangled Bell states, respectively.
In the case of a $\ket{\Psi^+}$ state, associated with an even wave-function parity (bosonic symmetry for the polarization observable), the two photons can always be related to an even spatial distribution due to their bosonic character. As a consequence, photon coalescence can be observed upon the projection by a PBS preceded by a 45$^{\circ}$ rotation of the bi-photon state compared to the $\{H,V\}$ basis, as demonstrated above.

\begin{figure}[h!]
\includegraphics[width=0.7\columnwidth]{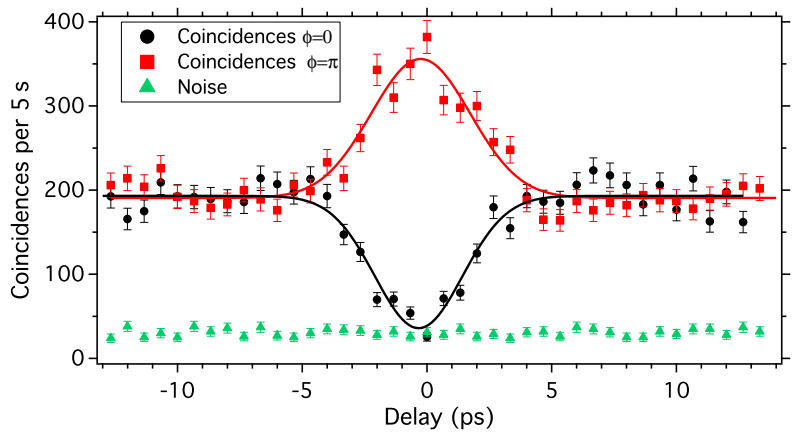}
\caption{HOM dip and peak obtained for two different SB settings. For $\phi=0$ the $\ket{\Psi^+}$ still holds and a dip is obtained, as was the case in \figurename~\ref{Fig5_HOMdipPhase}. Changing the phase to $\pi$ turns the state to $\ket{\Psi^-}$ for which the spatial distribution is expected to have an odd symmetry, therefore leading to a coincidence peak.\label{Fig5_HOMdipPhase}}
\end{figure}

If, however, a $\ket{\Psi^-} = \frac{1}{\sqrt{2}} \left( \ket{H}_{+}\ket{V}_{-}-\ket{V}_{+}\ket{H}_{-} \right)$ state, showing an odd wave-function parity (fermionic symmetry), is analyzed with such a setup, then the spatial distribution needs to be associated with an odd parity as well. In this case, after the bi-photon state is rotated by 45$^\circ$ using a fiber PC compared to the $\{H,V\}$ creation basis, projection at an f-PBS oriented in the $\{H,V\}$ basis leaves the state unchanged.
This is why an anti-coalescence behaviour is expected, \textit{i.e.} the paired photons are deterministically separated at the second f-PBS, therefore leading to increased coincidence rates.
\figurename~\ref{Fig3_Setup_DipPeak} depicts the experimental setup with the right inset (b) connected, which enables tuning the phase between the two contributions of the initial state $\ket{\Psi^+}$. After the polarization dependent delay line the paired photons are separated deterministically regarding their wavelengths using a first DWDM filter. In one arm, a variable delay is employed and in the other arm a phase shift between the contributions $\ket{H}_{+}$ and $\ket{V}_{+}$ is introduced thanks to a fiber coupled Soleil-Babinet (SB) phase plate. The paired photons are then recombined at a second DWDM, rotated by 45$^\circ$ using a fiber PC, and sent, eventually, to an f-PBS so as to perform HOM dip/peak measurements.
Note that this two-DWDM configuration, which provides deterministic separation and recombination of the photons, has the arrangement of a Mach-Zehnder interferometer. However, it requires no phase stabilization since the two arms are associated with a different wavelength, therefore preventing single-photon interference.
As shown in \figurename~\ref{Fig5_HOMdipPhase}, we observe again a clear dip when the phase introduced by the SB is $\phi=0$ and the birefringence compensation is made. If the phase is chosen to be $\pi$, then the maximally entangled state $\ket{\Psi^-}$ is generated and a coincidence peak is observed.
With the same detectors as for \figurename~\ref{Fig4_HOMdips}, both the dip and peak are associated with raw and net (\textit{i.e.} after subtraction of events associated with the dark-counts in the detectors) visibilities of 86$\pm$5\% and 94$\pm$6\%, respectively. Note that the curves of  \figurename~\ref{Fig5_HOMdipPhase} have been recorded for the same integration time as in \figurename~\ref{Fig4_HOMdips}. The reduced raw visibility is ascribed to an increase of the pump power to compensate the extra losses introduced in the setup of \figurename~\ref{Fig3_Setup_DipPeak}(b), since a free-space S/B and additional fiber coupling stages are used.

It is also remarkable that this behaviour is observed independently of the delay introduced between the $+$ and $-$ components. For the peak measurement of \figurename~\ref{Fig5_HOMdipPhase}, the fiber delay was of 22\,ns ($\leftrightarrow$5\,m), which is much greater than the $\sim$5\,ps ($\leftrightarrow$100\,GHz) single photons' coherence time travelling through the device. This condition therefore prevents them from temporally overlapping at the second f-PBS. As a conclusion of this analysis, which is similar to energy-time entanglement in a folded Franson configuration~\cite{Franson_Bell_1989,Thew_NonMax_2002}, the two properties discussed above can be attributed only to entangled states.

\section{Entanglement quality measurement}

\begin{figure}[h!]
\includegraphics[width=0.7\columnwidth]{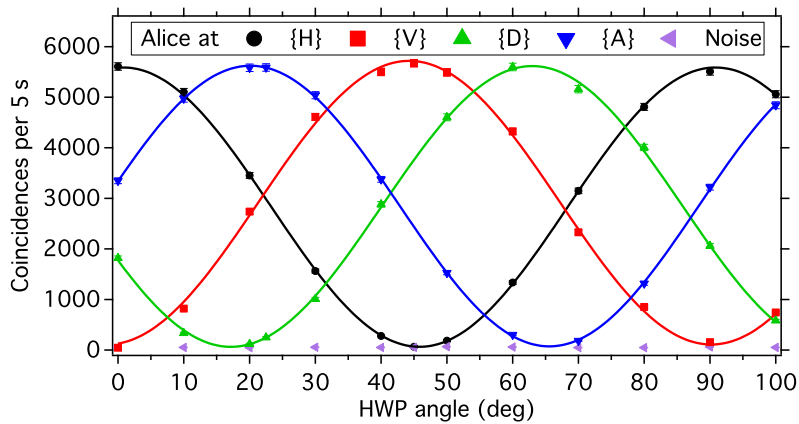}
\caption{Violation of the B-CHSH inequalities for the standard settings, \textit{i.e.} horizontal $\{H\}$, vertical $\{V\}$, diagonal $\{D\}$, anti-diagonal $\{A\}$. For all the settings, visibilities exceeding 99\% (net) and 97\% (raw) are obtained, which underlines the quality of our approach.\label{Fig6_Bell}}
\end{figure}

To further quantify the generated entanglement quality using the simple setup in \figurename~\ref{Fig1_Fullsetup}, we perform a Bell inequality test using the standard polarization settings~\cite{Zhong_typeIIsecond_2010,Martin_typeIIsecond_2010}. As discussed above, the birefringence compensation is made using a 3.2\,m long PMF. Then, the paired photons are deterministically split up and sent to Alice and Bob using two standard 100\,GHz DWDMs centred at the channels ITU-46 and ITU-47 (see \figurename~\ref{Fig2_Spectrum}). As the reflection of each DWDM is the complement of it's transmission window, two complementary filters are necessary. The generated entangled state is then analyzed by Alice and Bob, each of them comprising a fiber PC, a half-wave plate (HWP), a polarization beam-splitter (PBS), and a single photon detector. In order to reduce the noise in this measurement, we take advantage of two free-running InGaAs APDs (IDQ-220) featuring $\sim$20\% detection efficiency and $\sim$10$^{-6}$\,/\,ns dark-count probability. In addition, on Alice's side, an SB phase compensator is employed to adjust the phase relation $\phi$ between the contributions to the entangled state of \equationname{~\ref{Eq_Psi}}.
For the following measurements, we set $\phi=0$ for generating the maximally entangled Bell state $\ket{\Psi^+} = \ket{\Psi(0)}$. Then Alice fixes her polarization analyser to be subsequently oriented at horizontal $(H)$, vertical $(V)$, diagonal $(D)$ and anti-diagonal $(A)$. Simultaneously, the coincidence detection rate between Alice and Bob is measured as a function of Bob's HWP angle, which is continuously rotated. As shown in \figurename~\ref{Fig6_Bell}, we obtain excellent visibility interference patterns for all settings. The fitting parameters infer average raw and net (\textit{i.e.} after subtraction of events associated with the dark-counts in the detectors) visibilities of about 97.3$\pm$0.6\% and 99.5$\pm$0.8\%, respectively. These visibility figures correspond to net and raw Bell parameters of $S_{net}$=2.824$\pm$0.007 and $S_{raw}$=2.806$\pm$0.005, respectively, and therefore to the violation of the Bell, Clauser, Horne, Shimony, and Holt (B-CHSH) inequalities by more than 100 standard deviations for each value~\cite{CHSH_inequality_1969}. Note that the raw visibility is only limited by the dark-count probabilities in the detectors, in our case $\sim 1 \cdot 10^{-6}$\,/\,ns. The obtained net visibility underlines the high quality entanglement generated by the source itself. Note that similar results have been obtained for the $\ket{\Psi^-}$ state (curves not represented).

Regarding the other figures of merit of this source, we have reached a coincidence rate of about $1100\,\rm s^{-1}$ for a waveguide coupled pump power of 2.5\,mW. The spectral brightness of the PPLN/W was measured to be $\sim 2\cdot10^4$ generated pairs per second, per mW of coupled pump power, and per GHz of spectral bandwidth. In addition, the losses experienced by the single photons are of about 3\,dB, from the output of the generator to the measurement apparatus. This is an exceptionally good loss figure, and mainly ascribed to the advantageous use of off-the-shelf standard telecom components. Note that the main loss ($\sim 2$\,dB) comes from the butt-coupling between PPLN/W and the PMF, which could be strongly reduced using a fiber pigtail and/or a segmented tapered waveguide structure for enabling a better channel waveguide to fiber mode overlap~\cite{Castaldini_SPEtapers_2007}.

\section{Conclusion \& perspectives}

As a conclusion, we have demonstrated a high quality polarization entanglement source based on a type-II PPLN/W. The source is reliable and efficient thanks to the use of low-loss standard telecom components, such as a polarization maintaining fiber for waveguide birefringence compensation and a set of DWDMs for deterministic pair separation into standard communication channels. The excellent quality entanglement obtained at the output of the source (events due to the dark-counts in the detectors discarded), combined with the simplicity of the setup, make this approach a good candidate for future quantum networking solutions based on photonic entanglement. For instance, the reported source could be used in long-distance quantum key distribution schemes, provided detectors featuring much lower dark-count rates would be employed. Notably, state-of-the-art superconducting devices show dark-count probabilities on the order of 10$^{-9}$\,/\,ns~\cite{Verevkin_ultrafast_2004,Engel_Supra_2004}.

\section*{Acknowledgement}

The authors thank O. Alibart, A. Kastberg, M. P. De Micheli and  D.~B. Ostrowsky for their help and fruitful discussions. Financial support from the CNRS, l'Universit\'e de Nice -- Sophia Antipolis (UNS), both the European ERA-SPOT program ``WASPS'' and the ICT-2009.8.0 FET project ``QUANTIP'' (grant 244026), l'Agence Nationale de la Recherche project ``e-QUANET'' (grant ANR-09-BLAN-0333-01), le Minist\`{e}re de l'Enseignement Sup\'{e}rieur et de la Recherche (MESR), la Direction G\'en\'erale de l'Armement (DGA), the Malaysian government (MARA), la Fondation iXCore pour la Recherche, and le Conseil R\'{e}gional PACA, is acknowledged.

\end{document}